\documentstyle[11pt]{article}

\def\gt{{g^{(10)}}}
\def\gbt{{\bar{g}}^{(10)}}
\def\gb{{\bar{g}}}
\def\ght{{\hat{g}}^{(10)}}
\def\gh{{\hat{g}}}
\def\Sb{{\bar{S}}}
\def\Tb{{\bar{T}}}
\def\Phib{{\bar{\Phi}}}

\begin{document}
\begin{titlepage}
\vskip 3cm
\title{{\bf Dilaton, moduli and string/five-brane duality as seen from
four dimensions.}}
{\bf
\author{
        P.Bin\'etruy\thanks{ BINETRUY@FRCPN11.IN2P3.FR}\\
        Institute for Theoretical Physics\\
        University of California, Santa Barbara, Ca 93106\\
        and\\
        Laboratoire de Physique Th\'eorique et Hautes Energies
        \thanks{Laboratoire associ\'e au Centre National de la Recherche
        Scientifique.}\\
        Universit\'e de Paris-Sud, Bat. 211, 91405 Orsay, France.}
   }
\date{}
\maketitle
\vspace{-7cm}
\hfill NSF-ITP-93-60
\vspace{9cm}
\begin{abstract}
\noindent
A naive dimensional reduction of the $N=1, D=10$ supergravity theory
that naturally arises in five-brane models
is used to determine the r\^ole of two fields which are basic ingredients of
string models: the dilaton and, among the moduli, the breathing mode. It is
shown that, under the duality transformation that relates five-branes and
strings, these two fields exchange the r\^oles of 10-dimensional dilaton and
radius of the compact manifold. A description of this phenomenon in terms
of the linear multiplets of the 4-dimensional supergravity is also
presented.

\end{abstract}
\end{titlepage}

\newcounter{oldequation}

\def\subequations{\setcounter{oldequation}{\c@equation}%
  \setcounter{equation}{1}%
  \global\let\oldtheequation\theequation%
  \@ifundefined{chapter}%
       {\gdef\theequation{\theoldequation\alph{equation}}
        \global\let\thewholequation\theoldequation
        \gdef\@currentlabel{\theoldequation\alph{equation}}}
       {\gdef\theequation{\thechapter.\theoldequation\alph{equation}}
        \gdef\thewholequation{\thechapter.\theoldequation}
        \gdef\@currentlabel{\thechapter.\theoldequation\alph{equation}}}
  \global\let\@currenteqlabel\thewholequation}

\def\nosubequations{\setcounter{equation}{\c@oldequation}%
  \stepcounter{equation}
  \global\let\theequation\oldtheequation
  \global\let\@currentlabel\theequation}

\def\newsubequation{\stepcounter{oldequation}
    \setcounter{equation}{1}}

\def\@currenteqlabel{}

\def\eqlabel#1{\@bsphack\if@filesw {\let\thepage\relax
   \xdef\@gtempa{\write\@auxout{\string
      \newlabel{#1}{{\@currenteqlabel}{\thepage}}}}}\@gtempa
   \if@nobreak \ifvmode\nobreak\fi\fi\fi\@esphack}

\section{Introduction.}

The growing body of circumstantial evidence that superstrings are dual to
five-branes \cite{duff,andy,DL,CHS,DDP} in the critical dimension $d=10$
has led to think that one might tackle some of the problems raised in the
string formulation by going
over to the dual formulation. It was in particular suggested by
Strominger \cite{andy} that the weakly coupled five-brane is a dual
representation of the strongly  coupled heterotic string.

Although a heterotic five-brane was constructed by Strominger as a soliton
solution of the low-energy heterotic string field equations\cite{andy},
to this date there is no
complete covariant construction of the heterotic five-brane at the
quantized level. If such a theory can be constructed, we know that it would
yield a supergravity theory in $d=10$ dimensions in the formulation with a
seven-form field strength \cite{cham}. We can use this knowledge to
extract some information on the properties  of the fields involved at the
four-dimensional level.

One particular sector where new input is most eagerly awaited consists of
the dilaton and of the moduli fields. This set of fields, together with
their supersymmetric partners, undoubtedly plays a key r\^ole in the vast
problems of supersymmetry breaking, vanishing cosmological constant and,
probably, strong CP. Any information on the behavior of these fields
in the strongly interacting string regime is precious.
In the weakly interacting regime, it turns out that the bulk of
their properties
was extracted by Witten \cite{witty} by using a naive dimensional
reduction of 10-dimensional supergravity (in the formulation with a
three-form field strength \cite{CM}, dual to the one mentionned above)
mimicking  a
compactification on a Calabi-Yau manifold. More than  eight years later,
and although a lot of information has been gathered on string models,
this simple scheme of compactification  still adequately describes the
gross features of the dilaton and moduli properties.

In the case of the five-brane where a quantized version remains to be
constructed, there is no other way open presently
than to try a naive dimensional
reduction. The duality between the two theories lets us hope that more
information can be extracted this way than could be expected. In section
2, we undertake such a dimensional reduction, keeping in mind the
string version and the fact that the fields that we consider are
conjectured to be the same dilaton and moduli, only seen from the vantage
point
of a different string regime. We find that their r\^ole is actually exchanged
when one goes to the dual formulation: one of the moduli takes over as the
10-dimensional dilaton whereas the dilaton of the string picture plays the
r\^ole of the radius of the 6-dimensional compact manifold. This can be put
in perspective with an earlier observation \cite{DL} that the string/five-brane
duality interchanges the r\^oles of the $\sigma$-model loop expansion and
of the quantum loop expansion.

In section 3, we discuss this duality from the point of view of
4-dimensional supergravity. We stress that moduli, being associated in the
five-brane picture with antisymmetric tensors, are described by linear
multiplets and we discuss the relevance of some properties of the
geometrical structures associated with linear multiplets to some of the
issues at stake here. Finally, section 4 discusses the possible
implications of these results to the physics of the string dilaton and
moduli.

\vskip 1cm

\section{Dilaton and moduli in Planck, string and five-brane units.}

\vskip .5cm

We start by recalling the situation in the string case which will
constitute the backdrop of our discussion of the five-brane regime.
The action describing the low-energy field theory limit of the heterotic
string reads \cite{CFMP}, if we specify to the boson fields:
\begin{eqnarray}
S = {1 \over 2} \int d^{10}x \sqrt{-\gt}  \left( R^{(10)}
-{1\over 2} (\partial \phi)^2 - {1 \over 12} e^{-\phi} H^2
- {1 \over 4} e^{-\phi/2} F^2 + \ldots \right), \label{eq:st-can}
\end{eqnarray}
in agreement with 10-dimensional supergravity \cite{CM} (we have set the
Planck scale $M_{Pl}$ to 1).
In eq.(\ref{eq:st-can}), $R^{(10)}$ is the 10-dimensional curvature, $\phi$
is the dilaton, $F^{MN}$ is the Yang-Mills field strength and
$H_{MNP}= 3 \nabla_{[M} B_{NP]}$ is the field strength of the
antisymmetric tensor $B_{MN}$ which naturally appears at
the $\sigma$-model level as a background field, through the term:
\begin{equation}
{\cal S} = - {1 \over 2 \pi \alpha'} \int d^2\sigma \,\, {1 \over 2}
\epsilon^{ab} \partial_a X^M \partial_b X^N B_{MN}. \label{eq:st-2d}
\end{equation}

The simple dimensional reduction adopted by Witten in \cite{witty} amounts
to choose for the ten-dimensional metric (in the following, greek letters
$\mu,\nu,\ldots$ refer to 4-dimensional Lorentz indices while latin
letters $I,J,\ldots$ refer to compact indices):\footnote{The second of
these transformation laws ensures that $M_{Pl}$ is the same in 4 and 10
dimensions.}
\begin{eqnarray}
\gt_{IJ} &=& e^{\sigma} g^{(0)}_{IJ} \cr
\gt_{\mu\nu} &=& e^{-3\sigma} g_{\mu\nu},  \label{eq:dr-can}
\end{eqnarray}
where $\int d^{6}y \sqrt{g^{(0)}} = M_{Pl}^{-6}$. The field $e^{\sigma}$
plays the r\^ole of the radius of the compact manifold and $\sigma$ is
referred to as the breathing mode, the simplest example of a modulus.
Correspondingly, in order to remain compatible with four-dimensional
supersymmetry, one single massless mode is extracted from the part of the
antisymmetric tensor with compact indices:
\begin{equation}
B_{IJ} = \epsilon_{IJ} a_2(x), \hspace{2cm} (I,J) \in \{(4,5),(6,7),(8,9)\}
\label{eq:B}
\end{equation}

The dimensional reduction of the action (\ref{eq:st-can}) is easily seen to
be:
\begin{eqnarray}
S = {1 \over 2} \int d^4x \sqrt{-g} && \left[  R - {1 \over 2 s^2} \partial^\mu
s \partial_\mu s - {3 \over 2 t^2} \partial^\mu t \partial_\mu t \right. \cr
&& \left. \hspace{.3cm}
- {3 \over 2 t^2} \partial^\mu a_2 \partial_\mu a_2
-{1 \over 12} s^2 H^2 - {1 \over 4} s F^2 \right]  \label{eq:can-st}
\end{eqnarray}
with
\begin{eqnarray}
s = e^{-\phi/2} e^{3\sigma},  \hspace{1cm}   t = e^{\phi/2} e^{\sigma}.
\label{eq:s-t,can}
\end{eqnarray}
Eq.(\ref{eq:can-st}) is easily seen to agree with the standard formulation of
4-dimensional supergravity \cite{cremmer} by performing a duality
transformation on the antisymmetric tensor :
\begin{equation}
s^2 \,\, H^{\mu\nu\rho} = \epsilon^{\mu\nu\rho\sigma} \partial_\sigma a_1,
\label{eq:Ha1}
\end{equation}
and using a K\"ahler potential $K(S,\bar{S}) = - \ln (S+\bar{S})
- 3 \ln (T+\bar{T})$, where the complex scalar fields $S$ and $T$ are
defined as $S=s+ia_1$ and $T=t+ia_2$. But the natural formulation is in
terms of a linear supermultiplet \cite{FWZ,sante,BGGM2}
for describing both s and the invariant field
strength $H_{\mu\nu\rho}$ plus a chiral supermultiplet for $T=t+ia_2$. We
will come back to this point in the next section.

It is however somewhat more apropriate for discussing the r\^ole played by
each of these  fields to carry out the dimensional reduction
in string units, i.e.
to set $M_S=1$ instead of $M_{Pl} = 1$ \cite{DS}. Using the metric that
naturally arises in the string sigma-model, the 10-dimensional action
reads:
\begin{eqnarray}
S = {1 \over 2} \int d^{10}x \sqrt{-\gbt}  e^{-2\phi} \left( {\bar{R}}^{(10)}
+ 4 (\partial \phi)^2 - {1 \over 12} H^2
- {1 \over 4} F^2 + \ldots \right). \label{eq:st-sig}
\end{eqnarray}
The 10-dimensional dilaton thus naturally plays the r\^ole of the string
loop expansion (the $\ell^{th}$ order being proportional to
$e^{2(\ell-1)\phi}$).

In these string units, the dimensional reduction is even simpler than
previously \cite{DS}. Writing
\begin{eqnarray}
\gbt_{IJ} &=& e^{\sigma} \gb^{(0)}_{IJ} \cr
\gbt_{\mu\nu} &=& \gb_{\mu\nu}, \label{eq:dr-st}
\end{eqnarray}
where $\int d^{6}y \sqrt{\gb^{(0)}} = M_S^{-6}$, one obtains for the
4-dimensional action in natural string units
\begin{eqnarray}
S = {1 \over 2} \int d^4x \sqrt{-\gb} &s& \left[
\bar{R} - {1 \over 2 s^2} \partial^\mu
s \partial_\mu s - {3 \over 2 t^2} \partial^\mu t \partial_\mu t \right. \cr
&& \left. \hspace{.3cm}
- {3 \over 2 t^2} \partial^\mu a_2 \partial_\mu a_2
-{1 \over 12} H^2 - {1 \over 4} F^2 \right],  \label{eq:st}
\end{eqnarray}
with this time ( compare with eq.(\ref{eq:s-t,can}) )
\begin{eqnarray}
s = e^{-2\phi} e^{3\sigma},  \hspace{1cm}   t = e^{\sigma}.
\label{eq:s-t,st}
\end{eqnarray}

The two 4-dimensional actions (\ref{eq:can-st}) and (\ref{eq:st})
are of course in complete agreement. One is
obtained from the other by a Weyl rescaling:
\begin{equation}
g_{\mu\nu} = s \gb_{\mu\nu}   \label{eq:weyl-st}
\end{equation}
But the natural string units of eqs.(\ref{eq:st},\ref{eq:s-t,st}) make it
more transparent that $t$ is the radius of the compact manifold whereas $s$
retains all the properties of the 10-dimensional dilaton and is the string
loop expansion parameter as seen from 4 dimensions.

We now turn our attention to five-branes.\footnote{ We use here
and throughout this paper the term ``five-brane'' in
a loose sense since, strictly speaking, we are only dealing with 10-dimensional
supergravity in the formulation which is the dual of the one that
appears as the field theory limit of string theory.}
 In a canonical metric, the action
for the 10-dimensional effective field theory reads \cite{DL}
\begin{eqnarray}
S = {1 \over 2} \int d^{10}x \sqrt{-\gt}  \left( R^{(10)}
-{1\over 2} (\partial \phi)^2 - {1 \over 2\cdot 7!} e^{\phi} K^2
 + \ldots \right), \label{eq:fb-can}
\end{eqnarray}
where we have discarded for the moment Yang-Mills terms. In
(\ref{eq:fb-can}), $K = dA$ is the curl of the 6-form which naturally
appears as a background field at the sigma model level:
\begin{equation}
{\cal S} = - {1 \over 2 \pi \beta'} \int d^6\sigma {1 \over 6!}
\epsilon^{abcdef} \partial_a X^M \partial_b X^N \partial_c X^P \partial_d
X^Q \partial_e X^R \partial_f X^S A_{MNPQRS}. \label{eq:fb-2d}
\end{equation}
The action (\ref{eq:fb-can}) is simply the dual version of the string action
(\ref{eq:st-can}) if we interpret the 7-form as the dual of the 3-form
encountered previously:
\begin{equation}
K = e^{-\phi} {}^*H.
\end{equation}
Seen from the 4-dimensional point of view, this has the interesting
consequence that the moduli are now connected through supersymmetry with an
antisymmetric tensor and therefore described by linear multiplets whereas
the dilaton now fits into a chiral supermultiplet. In order to see that, we
will perform the same dimensional reduction as in the string case, that is
using eq.(\ref{eq:dr-can}). A general dimensional reduction of the
10-dimensional action has been performed in Ref.\cite{cham}. We define
\begin{eqnarray}
K_{\mu \nu \rho IJKL} & \simeq & {\cal K}_{\mu \nu \rho}, \cr
K_{\mu IJKLMN} & \simeq & \partial_\mu a_1,  \label{eq:Kdef}
\end{eqnarray}
where we have restricted our attention to the case of a single 3-form in
order to parallel exactly the previous discussion ( a single modulus was
considered in eq.(\ref{eq:B})). Then we obtain for the
4-dimensional action:
\begin{eqnarray}
S = {1 \over 2} \int d^4x \sqrt{-g} && \left[  R - {1 \over 2 s^2} \partial^\mu
s \partial_\mu s - {3 \over 2 t^2} \partial^\mu t \partial_\mu t \right. \cr
&& \left. \hspace{.3cm}
- {1 \over 2 s^2} \partial^\mu a_1 \partial_\mu a_1
-{1 \over 12} t^2 {\cal K}^2  \right],  \label{eq:can-fb}
\end{eqnarray}
with $s$ and $t$ still given by eq.(\ref{eq:s-t,can}). Performing two
inverse duality transformations, namely eq.(\ref{eq:Ha1}) and
\begin{equation}
t^2 \,\, {\cal K}_{\mu\nu\rho} = \sqrt{3} \epsilon_{\mu\nu\rho\sigma}
\partial^{\sigma}a_2
\label{eq:Ka2}
\end{equation}
we find the same action as in eq.(\ref{eq:can-st}). In other words, the
theory, to this order is still described {\it on-shell} \footnote{Since we
use duality transformations.} by the same K\"ahler potential
$K(S,\bar{S}) = - \ln (S+\bar{S}) - 3 \ln (T+\bar{T})$.

However, just as in the string case, we expect to get more meaningful
information by going over to natural five-brane units ($M_B = 1$).
Indeed, using the
metric which appears naturally in the five-brane sigma model, the
10-dimensional action now reads:
\begin{eqnarray}
S = {1 \over 2} \int d^{10}x \sqrt{-\ght}  e^{2\phi/3} \left( {\hat{R}}^{(10)}
- {1 \over 2\cdot 7!} K^2 + \ldots \right).
\label{eq:fb-sig}
\end{eqnarray}
The loop expansion parameter is now given by $e^{2\phi/3}$.
It has been noted by Duff and Lu \cite{DL} that, surprisingly enough, the
dilaton field has no kinetic term in this frame. Of course, it is still a
propagating degree of freedom as can be seen from the Einstein frame in
eq.(\ref{eq:fb-can}) but it is interesting to note that a similar phenomenon
occurs in four dimensions: in no-scale models, the dilaton field associated
with the flat directions has a vanishing kinetic term in a specific frame
\cite{EKN}. We are going to see that the comparison is not fortuitous.

Indeed the compactification goes as follows in five-brane units. We write
(compare with eq. (\ref{eq:dr-st})):
\begin{eqnarray}
\ght_{IJ} &=& e^{\sigma} \gh^{(0)}_{IJ} \cr
\ght_{\mu\nu} &=& e^{-2\sigma} \gh_{\mu\nu}, \label{eq:dr-fb}
\end{eqnarray}
with $\int d^6y \sqrt{\gh^{(0)}} = M_B^{-6}$, and use the definitions
(\ref{eq:Kdef}) to obtain the dimensionally reduced action:
\begin{eqnarray}
S = {1 \over 2} \int d^4x \sqrt{-\gh} &t& \left[  \hat{R}
- {1 \over 2 s^2} \partial^\mu s \partial_\mu s
 \right. \cr
&& \left. \hspace{.3cm}
- {1 \over 2 s^2} \partial^\mu a_1 \partial_\mu a_1
-{1 \over 12} t^2 {\cal K}^2  \right],  \label{eq:fb}
\end{eqnarray}
where now $s$ and $t$ are given in five-brane units by (compare with
eq.(\ref{eq:s-t,st})):
\begin{eqnarray}
s = e^{3\sigma},  \hspace{1cm}   t = e^{\sigma} e^{2\phi/3}.
\label{eq:s-t,fb}
\end{eqnarray}

It is straightforward to show that such an action can also be derived from
the action (\ref{eq:can-fb}) through the rescaling
\begin{equation}
g_{\mu\nu} = t \gh_{\mu\nu}.  \label{eq:weyl-fb}
\end{equation}
Nevertheless, the derivation in five-brane units sheds new light on several
aspects. First, one checks that the $t$ fields has no kinetic term which is
related to the fact that it is the scalar field connected with the no-scale
structure of the theory (the factor 3 in the logarithmic term of the
K\"ahler potential). Moreover, eq.(\ref{eq:s-t,fb}) clearly shows that the
$t$ field is the one that retains the character of the 10-dimensional
dilaton, as well as the five-brane loop expansion parameter. On the other hand,
the $s$ field is now the breathing mode associated with
the compactification of the five-brane theory. It is also the five-brane
sigma model coupling.\footnote{ From eq.(\ref{eq:dr-fb}), it seems that,
both in the limit $s \rightarrow 0$ and $s \rightarrow \infty$, the
$\sigma$-model is strongly coupled.}
In other words, compared
with the string case, the r\^ole of the two scalar fields is interchanged
\cite{DL}. Thus  the duality $\phi \rightarrow -\phi/3$ between the string
action (\ref{eq:st-sig}) and the five-brane action (\ref{eq:fb-sig}) does not
seem to be related to a $S \rightarrow 1/S$ ``duality'' \cite{FILQ} but rather
to an exchange of the r\^oles of $S$ and $T$. On the other hand, in the
five-brane regime, since $s$ is interpreted as the radius, a $S \rightarrow
1/S$ ``duality'' --~or more precisely an $SL(2,Z)$ invariance~-- does  seem
plausible (whereas the $SL(2,Z)$ invariance associated with the $T$ field
in the string formulation is no longer explicit in the five-brane picture).

It is interesting to note that adding in the action ({\ref{eq:fb-sig}) a
gauge term of the form:
\begin{equation}
\delta S = {1 \over 2} \int d^{10}x \sqrt{-\ght} \left( - {1 \over 4 } F^2
\right)  \label{eq:gauge10}
\end{equation}
yields in four dimensions
\begin{eqnarray}
\delta S&=&{1/2} \int d^4x \sqrt{-\gh} \left( -{1\over4} s F^2 \right) \cr
&=&{1/2} \int d^4x \sqrt{-g} \left( -{1\over4} s F^2 \right),
\label{eq:gauge4}
\end{eqnarray}
{\it i.e.} a standard gauge term in 4 dimensions (see
eq.(\ref{eq:can-st})). The absence of a factor $e^{2\phi/3}$ in the
10-dimensional Lagrangian (or a factor $t$ in the 4-dimensional Lagrangian
in five-brane units) shows that such a term is only expected at the
one-loop level. e will return to this in the next section.

\section{Linear multiplets and duality transformations.}

In $N=1$  supergravity in 4 dimensions, the natural framework to deal with an
antisymmetric tensor field is the linear supermultiplet \cite{FWZ,sante,BGGM2}.
In the weakly
coupled string regime, this has proven to be useful to describe the
couplings of the dilaton $s$, the antisymmetric tensor field strength
$H_{\mu\nu\rho}$  and their fermionic partner \cite{CFV,ABGG,DFKZ,GT}.
The moduli, on the other hand, are described by chiral supermultiplets
since there is no four-dimensional antisymmetric tensor involved. One
should note however that their pseudoscalar partners are provided by some
of the components of the ten-dimensional antisymmetric tensor (see for
example eq.(\ref{eq:B})).

In the five-brane regime, the situation is reversed and the $s$ field
fits into a chiral supermultiplet whereas the moduli are now described by
linear multiplets. The geometrical structure of multi-linear  supermultiplets
has some interesting  features \cite{BGGprep} that we will sketch below. But
we will start by describing the duality transformations at the supergravity
level.

In fact, in order to show the specificity of the duality transformation
involved when going from the string regime to the five-brane regime, we will
consider for the time being a slightly more general situation.

Consider a theory with a linear multiplet $L_1$, a chiral supermultiplet
$T$ and some matter chiral supermultiplets which we will denote
generically by $\Phi$,
whose interactions are described by the action (we work in the K\"ahler
superspace of Refs.\cite{BGGM1,BGGM2,BGG})\footnote{ In this section,
$\int$ stands for $\int d^4x d^4\theta$.}
\begin{equation}
S = -3 \int E \left[ a  + L_1 V(\Phi, \Phib) \right]  \label{eq:S}
\end{equation}
and a K\"ahler potential:
\begin{equation}
K = -\alpha \ln ( T + \Tb + W(\Phi,\Phib)) + \beta \ln (L_1). \label{eq:K}
\end{equation}
In our case, $\alpha = 3$ and $\beta=1$ but, as said above, we will keep
them general for the time being (assuming $\beta \neq 3$). Also in
eq.(\ref{eq:S}), $E$ is the supervierbein determinant, $a$ is a real constant
and $V(\Phi, \Phib), W(\Phi, \Phib) $
are general real functions of the matter fields. Indeed, had we included a
$T$ dependence in $V$, eq.(\ref{eq:S}) would be
the most general Lagrangian that we can write since a Weyl rescaling can
absorb terms of order $L_1^n,\; n \neq 1$ in the K\"ahler potential ($E$ has
Weyl weight $-2$ and $L_1$ has Weyl weight $2$). Of course, eq.(\ref{eq:K})
{\it is not} the most general K\"ahler potential that one can write. Also,
we did not write the superpotential terms but we are assuming here that $T$
does not appear in them.

We are now going to make a ``dual'' duality transformation which will
replace $L_1$ by the chiral superfield $S$ and $T$ by the linear multiplet
$L_2$. The method is standard and amounts to a Legendre transformation
\cite{LR,sante,BGGM2}}. One starts with the action:
\begin{equation}
S = -3 \int E \left[ a  + L_1 V(\Phi, \Phib) + L_1(S+\Sb) - L_2 Y\right]
\label{eq:Sduality}
\end{equation}
with a K\"ahler potential:
\begin{equation}
K = -\alpha \ln ( Y + W(\Phi,\Phib)) + \beta \ln (L_1), \label{eq:Kduality}
\end{equation}
$L_1$ and $Y$ being {\it unconstrained real superfields}, $L_2$ a linear
superfield and $S$ a chiral one.

If we minimize with respect to the constrained superfields $S$ and $L_2$,
we obtain:
\begin{eqnarray}
({\cal D}^\alpha {\cal D}_\alpha - 8 R^{\dagger})L_1 = 0 &,&
({\cal D}_{\dot{\alpha}} {\cal D}^{\dot{\alpha}} - 8 R) L_1 = 0  \cr
Y = T + \Tb &,& {\cal D}^{\dot{\alpha}} T =0, \label{eq:constraint}
\end{eqnarray}
and we therefore recover the previous theory described by the action
(\ref{eq:S}).

In order to get the dual theory, we can alternatively minimize with respect
to the unconstrained superfields $L_1$ and $Y$. We obtain
respectively:\footnote{ In K\"ahler superspace where the K\"ahler invariance is
implemented into the superspace structure, we have $\delta_{L_1} E =
-{1 \over 3} {\partial K \over \partial L_1} E \delta L_1$ and $\delta_Y E =
-{1 \over 3} {\partial K \over \partial Y} E \delta Y$ (see for example
ref.\cite{ABGG}). Similarly, because the constraint on $L_2$ involves the
K\"ahler connection, $\delta_{L_1} L_2 =
{1 \over 3} {\partial K \over \partial L_1} L_2 \delta L_1$ and $\delta_Y
L_2 = {1 \over 3} {\partial K \over \partial Y} L_2 \delta Y$ (on the other
hand, $S$ being of chiral weight zero, its constraint does not involve the
K\"ahler potential).}
\begin{eqnarray}
-{1 \over 3} {\partial K \over \partial L_1} \left[ a + L_1 ( S + \Sb + V )
\right] + S + \Sb + V  &=& 0, \cr
-{1 \over 3} {\partial K \over \partial Y} \left[ a + L_1 ( S + \Sb + V )
\right] - L_2  &=& 0.
\end{eqnarray}
Using the explicit form for the K\"ahler potential, eq.(\ref{eq:Kduality}),
we can easily express $L_1$ and $Y$ in terms of $S$ and $L_2$ and we
finally obtain the following action, dual to (\ref{eq:S}),
\begin{equation}
S = -3 \int E \left[ a \; {3-\alpha \over 3-\beta}
+ L_2 W(\Phi, \Phib) \right]
\label{eq:Sdual}
\end{equation}
with the K\"ahler potential:
\begin{equation}
K = \alpha \ln (L_2) - \beta \ln( S + \Sb + V(\Phi,\Phib))
- \alpha \ln \alpha + \beta \ln \beta +(\alpha - \beta) \ln {3-\beta \over
a}. \label{eq:Kdual}
\end{equation}

It is straightforward to check that this ``dual'' duality transformation
can be inverted, {\it provided $\alpha$ and $\beta$ are different from
$3$}.

It is very interesting indeed that the case of the string/five-brane
duality corresponds to a singular situation, {\it i.e.} $\alpha =3$. We can
see it from the fact that the action (\ref{eq:Sdual}) is in this case scale
invariant. Such an action has been developped in terms of component fields in
Ref.\cite{ABGG}, but, due to the singular properties of the limit $\alpha=3$,
the methods used there cannot be readily applied to this case. Moreover, as
we have just said, the duality transformation cannot be inverted to get
back to the string Lagrangian.

It might be that the ``bare bone'' Lagrangian of eq.(\ref{eq:S})
cannot, by itself, describe the low energy string Lagrangian if a dual
theory such as the five-brane can be constructed. We actually do know that
eq.(\ref{eq:S}) is not the complete story because the Green and Schwarz
anomaly-cancelling mechanism \cite{GS} has to be implemented at the
4-dimensional level.\footnote{ Note that this is a one-loop effect.}
It has actually been shown that this is most easily
done in the linear multiplet formulation \cite{CFV,DFKZ}. The term linear
in $L_1$ in (\ref{eq:S}) now depends on the $T$ field and we must couple
$L_1$ to the Chern-Simons form \cite{richard,sante,BGGM2} through the
constraint
\begin{eqnarray}
({\cal D}^\alpha {\cal D}_\alpha - 8 R^{\dagger})L_1 &= 2k \; tr ({\cal
W}_{\dot{\alpha}} {\cal W}^{\dot{\alpha}}) &\equiv k \;
({\cal D}_{\dot{\alpha}} {\cal D}^{\dot{\alpha}} - 8 R) \Omega, \cr
({\cal D}_{\dot{\alpha}} {\cal D}^{\dot{\alpha}} - 8 R) L_1 &=  2k \; tr ({\cal
W}^\alpha {\cal W}_\alpha) &\equiv k \; ({\cal D}^\alpha {\cal D}_\alpha
- 8 R^{\dagger}) \Omega, \label{eq:constraint'}
\end{eqnarray}
where we have restricted our attention to Yang-Mills Chern-Simons forms,
decribed by the superfield $\Omega$ ($k$ is a normalisation constant).

We thus start instead of (\ref{eq:Sduality}) with the action:
\begin{equation}
S = -3 \int E \left[ a  + L_1 V(Y, \Phi, \Phib) + (L_1- \Omega)(S+\Sb)
- L_2 Y\right]
\label{eq:Sduality'}
\end{equation}
where the K\"ahler potential is still given by eq.(\ref{eq:Kduality})
(with $\alpha = 3$ and $\beta = 1$) and we choose \cite{DFKZ}
\begin{equation}
V(Y, \Phi, \Phib) = c \ln ( Y + W( \Phi, \Phib)). \label{eq:V}
\end{equation}
Minimizing with respect to the constrained fields $L_2$ and $S$, one
recovers the constraints for $Y$ in eq.(\ref{eq:constraint}) and the
modified constraint of eq.(\ref{eq:constraint'}) for $L_1$.

On the other hand, minimizing with respect to the unconstrained fields, one
now obtains:
\begin{eqnarray}
S + \Sb &=& {a \over 2 L_1} - c \ln (Y+W) \cr
L_2 &=& {1 \over Y+W} \left( {3a \over 2} + cL_1 \right) \label{eq:min}
\end{eqnarray}
Thus the action reads:
\begin{eqnarray}
S &=& -3 \int E \left[ L_2 W(\Phi,\Phib) - c L_1(S+\Sb,L_2) \right] \cr
& & + 3 \int E \Omega (S +\Sb)  \label{eq:Sdual'}
\end{eqnarray}
where $L_1(S+\Sb,L_2)$ is the solution of the equation
\begin{equation}
{a \over 2 L_1} - \ln \left( {3a \over 2} + c L_1 \right) =
S + \Sb - c \ln L_2.  \label{eq:L_1}
\end{equation}
The second term in the action is nothing but the standard Yang-Mills
kinetic term \cite{ABGG} which thus appears as a one-loop effect in this
dual formulation (see the comments at the end of the previous section):
\begin{equation}
S = - {3\over4} k \int {E \over R} S Tr({\cal W}^\alpha {\cal W}_\alpha)
- {3\over4} k \int {E \over R^{\dagger}} \Sb Tr({\cal W}_{\dot{\alpha}}
{\cal W}^{\dot{\alpha}}).
\end{equation}
{}From eq.(\ref{eq:L_1}), we see that $L_1$ depends only on the combination
$S + \Sb - c \ln L_2$ which is nothing but the renormalised gauge coupling
of the string theory $1/g^2_{1loop}$. In the limit of large $c L_1$,
one obtains for the first term of the action
\begin{equation}
S = -3 \int E \left[ L_2 W(\Phi,\Phib) - e^{-(S+\Sb -c\ln L_2)} \right]
\end{equation}
where the second term is of order $e^{-1/g^2}$, whereas in the limit of
small $L_1$, one obtains
\begin{equation}
S = -3 \int E \left[ L_2 W(\Phi,\Phib) - {a \over 2}
{c \over S+\Sb -c\ln L_2} \right]
\end{equation}
where the second term is of order $g^2$.

Let us finally make a few comments about the more realistic situation where
one considers several moduli, and thus, in the dual formulation, several
linear multiplets.

There is an interesting geometrical structure of the
projective type involved with the superspace formulation of theories
with several linear multiplets $L^A, A=1 \ldots N$ \cite{BGGprep}. This is
already apparent at the level of the duality transformation \cite{ABGG}
which can be reformulated in terms of
one given $L^B$ and of the homogeneous variables $\xi^A_B \equiv L^A/L^B,
A \neq B$, which play a spectator role in the duality
transformation. The procedure does not depend on the choice of $L^B$
{\it i.e.} it is invariant under the transformations $\xi^A_B
\rightarrow \xi^A_C = \xi^B_C \xi^A_B$.
This might indeed be of relevance for our discussion above since it
is difficult to understand why the $T$ field should  play such a special
r\^ole in the dual formulation. In fact, this projective structure can be
implemented into the superspace structure by introducing derivatives
covariant under the above transformations; $\xi^A_B$ then satifies the
accordingly modified linear multiplet constraints \cite{BGGprep}.

\section{Conclusions.}

If the conjecture that weakly coupled five-brane theories form a dual
representation of strongly coupled heterotic strings is verified, this
might have deep implications for long-standing problems such as the breaking of
supersymmetry. Indeed both the dilaton and the moduli play an important
r\^ole in this problem. Seen from the $N=1$, $D=4$ supergravity point of view,
one might have to allow for more general couplings than the ones identified so
far. This was hinted at in the last section by considering the duality
transformation. Another example is the following: if moduli are components
of linear supermultiplets in the dual formulation, how can they appear  in
the superpotential? Also, regarding supersymmetry breaking, the possibility
of having a formulation where a $SL(2,Z)$ symmetry associated with the
``string coupling'' $S$ is manifest is certainly tantalizing \cite{FILQ}.

Actually, while completing this work, we realized that, in a recent
preprint \cite{new}, J.\ Schwarz and A.\ Sen have addressed some of the issues
discussed here, precisely in the spirit of formulating the theory in a
manifestly $SL(2,Z)$ invariant way. They perform a dimensional reduction of
the dual formulation of 10-dimensional supergravity which is more complete
than the one presented here where our main purpose was to
unravel the r\^ole of the different fields. We refer the reader to
their discussion of the $SL(2,Z)$ symmetry in the dual formulation which we
have only briefly mentionned here.

\vskip 2cm
{\bf Acknowledgments.}
\vskip 1cm
It is a pleasure to thank M.\ Duff, M.K.\ Gaillard, R.\ Grimm, J.\ Schwarz
and A.\ Sen for helpful remarks and discussions. This work was
supported in part by National Science Foundation under grant
no. PHY89-04035.

\vfill
\end{document}